\documentclass[reprint, aps, amsmath, amssymb, twocolumn]{revtex4-2}

\usepackage{dirtytalk}
\usepackage{graphicx}
\usepackage{dcolumn}
\usepackage{bm}

\usepackage[caption=false]{subfig}
\usepackage[utf8]{inputenc}
\usepackage[table]{xcolor}
\usepackage{physics}
\usepackage{amsmath}

\usepackage{url}
\usepackage[hidelinks]{hyperref}
\numberwithin{equation}{section}
\renewcommand\theequation{\arabic{section}.\arabic{equation}}

\begin{document}

\title{Transition to Turbulence in Driven Active Matter}

\author{Aritra Das}
\email{aritrab@iitk.ac.in}
\affiliation{Department of Physics, Indian Institute of Technology Kanpur, Kalyanpur 208016, Uttar Pradesh, India}

\author{J. K. Bhattacharjee}
\email{tpjkb@iacs.res.in}
\affiliation{Department of Theoretical Physics, Indian Association for the Cultivation of Science, 2A and 2B Raja S C Mullick Road, Kolkata 700032, West Bengal, India}

\author{T. R. Kirkpatrick}
\email{tedkirkp@umd.edu}
\affiliation{Institute for Physical Science and Technology, University of Maryland, College Park, Maryland 20742}

\date{\today}

\begin{abstract}
A Lorenz-like model was set up recently, to study the hydrodynamic instabilities in a driven active matter system. This Lorenz model differs from the standard one in that all three equations contain non-linear terms. The additional non-linear term comes from the active matter contribution to the stress tensor. In this work, we investigate the non-linear properties of this Lorenz model both analytically and numerically. The significant feature of the model is the passage to chaos through a complete set of period-doubling bifurcations above the Hopf point for inverse Schmidt numbers above a critical value. Interestingly enough, at these Schmidt numbers a strange attractor and stable fixed points coexist beyond the homoclinic point. At the Hopf point, the strange attractor disappears leaving a high-period periodic orbit. This periodic state becomes the expected limit cycle through a set of bifurcations and then undergoes a sequence of period-doubling bifurcations leading to the formation of a strange attractor. This is the first situation where a Lorenz-like model has shown a set of consecutive period-doubling bifurcations in a physically relevant transition to turbulence.
\end{abstract}

\maketitle

\section{Introduction}

The route to turbulence in a fluid as some control parameter is varied (e.g. increasing the mean flow velocity, or increasing the density gradient if the instabilities are in a stratified fluid) was first conjectured by Landau \cite{ref1}. In Landau's scenario, a steady motion becomes a periodic motion and then the periodic motion is destabilized to yield a quasi-periodic motion with two incommensurable frequencies and subsequently, the number of incommensurate frequencies increases with increasing control parameter and eventually the motion becomes a-periodic. This sequence of events was shown to be improbable by Ruelle, Takens, and Newhouse \cite{ref2, ref3} who argued that any more than three quasi-periodic frequencies would lead to an a-periodic state. In a parallel development, to study the a-periodic flows in a fluid stratified by heating from below, Lorenz \cite{ref4} decided to use a Galerkin truncation procedure to reduce the coupled (velocity and local temperature) partial differential equations to a set of ordinary differential equations. Because of the nature of the fluid (having viscosity and thermal diffusivity), the system had to be dissipative. The minimum number of modes in such a truncation would have to be three to bypass the Poincare-Bendixon theorem \cite{ref5}. Choosing the three most relevant modes Lorenz obtained a set of three coupled ordinary differential equations now known as the Lorenz model. The control parameter ($r$) of the model was the temperature gradient while there were two other system parameters. One of them ($b$) describes the geometry of the setup ($b = \frac{8}{3}$ for Lorenz) and the other is the Prandtl number of the fluid. The Lorenz model took the form 

\begin{subequations}
    \begin{equation}
        \dot{X} = \sigma(-X+Y) \label{lor1}
    \end{equation}
    \begin{equation}
        \dot{Y} = - Y + r X - XZ  \label{lor2}
    \end{equation}
    \begin{equation}
        \dot{Z} = XY - bZ  \label{lor3}
    \end{equation}
\end{subequations}
The modes $X$ and $Y$ stand for the convective roll patterns set up by the velocity and temperature fields and $Z$ is the amount of convective heat transfer across the fluid. For very small gradients ($r<1$)  the conduction state ($X = Y = Z = 0$) is stable. It becomes unstable and steady convection sets in for $r>1$. The steady-state is destabilised by a Hopf bifurcation at $r = r_c = \sigma \frac{\sigma + b + 3}{\sigma - b -1}$ but a periodic state is not observed for $r>r_c$ since the bifurcation is backward. Lorenz found that the a-periodic behavior sets in almost immediately above $r_c$ and further, the trajectories although confined in a region, are very sensitive to the initial conditions. The set of points to which all trajectories are attracted is called a strange attractor. The signature of fluid turbulence in the truncated model of Eqs. \eqref{lor1}---\eqref{lor3} was the extreme sensitivity of trajectories to initial conditions---a state of affairs described as chaos. Thus the Lorenz model described a new scenario in which the onset of turbulence followed the sequence: trivial steady-state $\to$ convective steady-state $\to$ unstable periodic state via a Hopf bifurcation $\to$ turbulence.

More than a decade later a new paradigm was discovered in the study of low dimensional return maps \cite{ref6}---\cite{ref8}. This was an infinite sequence of bifurcations leading to periodic orbits with higher and higher time periods. If the first Hopf bifurcation produces a periodic state of time period $T$, then the successive states are characterized by $2T, 2^2T, \dots , 2^n T, \dots $ with new periods appearing with only a slight change in the control parameter. Soon, the system becomes a-periodic at a critical value of the control parameter and also shows sensitive dependence on initial conditions. An expected pathway to turbulence could then be steady-state $\to$ periodic state of period $T$ $\to$ $2T$ $\to$ $2^2 T$ $\to$ $\dots$ $\to$ a-periodic state. This sequence, or more correctly, some part of it has been seen in some damped driven Duffing oscillators \cite{ref9}, but never in any physically motivated Lorenz-like model of a system governed by non-linear partial differential equations. In real experiments, one of the earliest observations of a few period doublings is that of Libchaber et al \cite{ref10}.

A slightly different kind of hydrodynamical problem is posed by ``active matter'' (bacteria swimming in a fluid) \cite{ref11}---\cite{ref13}. Here, because of its own energy source, active matter can exert an additional stress in the Navier-Stokes equation. The usual stress tensor for the velocity dynamics of an incompressible fluid is $T_{ij} = - p \delta_{ij} + \eta (v_{i, j} + v_{j, i})$ where $v_{i, j} = \partial_j v_i$, $p$ is the pressure and $\eta$ is the shear viscosity. The additional contribution to $T_{ij}$ can take different forms \cite{ref14}---\cite{ref17}, depending on what is being studied and a particular form \cite{ref18} of this extra term is reminiscent of model H in the different universality classes of dynamic critical phenomena \cite{ref19}. In Ref. \cite{ref20} the driven form of active matter was studied where a gradient in the density of active particles was considered in one particular direction. To study the effect of non-linearity in this problem a Lorenz-like model was set up. The crucial difference with the usual Lorenz model of Eqs. \eqref{lor1}---\eqref{lor3} is the existence of an extra non-linear term in Eqn. \eqref{lor1}. In Ref. \cite{ref20} we studied the properties of the steady-state and showed how they were significantly different from that of the standard Lorenz model.

\begin{figure}[h]
    \centering
    \includegraphics[width = 90 mm]{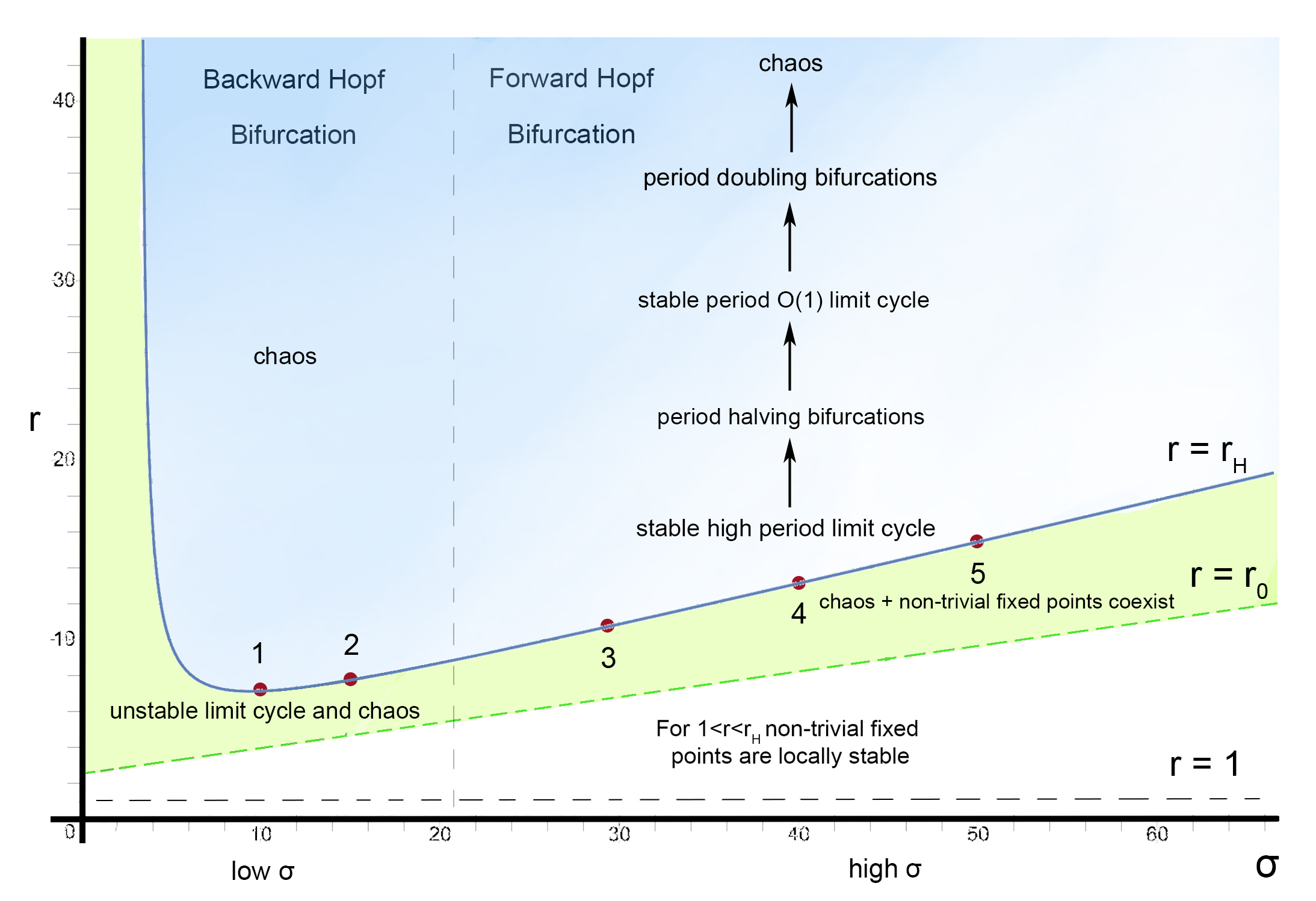}
    \caption{Summary of our results on an illustrated $r$ vs $\sigma$ plot.}
    \label{fig:finalplot}
\end{figure}

In this work, we concentrate on the nature of the Hopf bifurcation. We find that this version of the Lorenz model has the following series of bifurcations for small Schmidt numbers:
\\* Trivial steady-state $\to$ convective steady-state with two stable fixed points $\to$ a Hopf bifurcation to two symmetrically situated periodic orbits of period $T$ about the previously stable fixed points $\to$ a sequence of period-halving bifurcations leading to one periodic orbit encircling both the unstable fixed points $\to$ a sequence of period-doubling bifurcations leading to chaos.
\\* As far as we know, this is the first situation where a Lorenz-like model has shown a set of consecutive period-doubling bifurcations in a physically relevant scenario. A complete summary of our principal results is shown in Fig. \ref{fig:finalplot}.

The layout of the paper is as follows: In Sec. \ref{sec2} we study the convective instability of the trivial state in the governing partial differential equations and then set up the relevant Lorenz model and study the non-trivial steady state. In Sec. \ref{sec3}, the Hopf bifurcation is examined and the system reduces to the canonical form near the bifurcation point. The linear part is shown to lead to a growing mode as the control parameter is increased. The non-linear part shows a forward bifurcation for $\sigma \gg 1$ but a backward bifurcation for small $\sigma$. In the process, we extend the Krylov-Bogoliubov technique \cite{ref21, ref22} beyond its usual habitat of two-dimensional dynamical systems, the details of which are given in Appendix \ref{Appen}. In Sec. \ref{sec4} we present detailed numerical results for the model which show the existence of a complete sequence of period-doubling bifurcations leading to chaos as shown in Fig. \ref{fig:finalplot}. We examine closely the region $r_0 < r < r_c$, $r_0$ being the point where the homoclinic bifurcation occurs and $r_c$ a point slightly above the Hopf bifurcation point $r_H$, and find some rather striking features. We conclude with a summary in Sec. \ref{sec5}.

\section{The Lorenz Model for Active Matter}
\label{sec2}
We consider wet active matter in it's simplest form, by ignoring the orientational degrees of freedom. This model was proposed by Tiribocchi et al. \cite{ref18} and is the active fluid version of Model H in the Halperin-Hohenberg classification scheme.

The variables of interest are:
\begin{enumerate}
    \item The scalar concentration field $\phi(\bold{r}, t)$ of the active particles (also called swimmers) which is proportional to their density, and
    \item The vector velocity field $\bold{u}(\bold{r}, t)$ which represents the velocity of the fluid medium in which the active matter moves about.
\end{enumerate}
The fluid velocity follows an incompressible Navier-Stokes equation which takes the form

\begin{subequations}
\begin{align}
    \partial_t u_i + u_j \partial_j u_i &= - \partial_i p + \nu \nabla^2 u_i + \partial_j  \Sigma_{ij} \label{goveq2} \\
    \partial_i u_i &= 0 \label{goveq1}
\end{align}
\end{subequations}
\\*where, $p(\bold{r}, t)$ is the pressure field, $\nu$ is the kinematic viscosity and $\Sigma_{ij}$ is the stress tensor generated by the active particles. To the lowest non-trivial order allowed by symmetry considerations, it has the form 

\begin{equation}
    \Sigma_{ij} = -\zeta\left (\partial_i \phi \partial_j \phi - \frac{\delta_{ij}}{3} \left (\nabla \phi \right )^2 \right )
    \label{stresstensor}
\end{equation}
\\*where $\zeta$ is a constant which can be termed the activity coefficient. This additional term in the Navier-Stokes equation is the non-linear Burnett term. The concentration field $\phi(\bold{r}, t)$ satisfies the usual advection-diffusion equation:

\begin{equation}
    \partial_t \phi(\bold{r}, t) + u_j\partial_j \phi(\bold{r}, t) = D \nabla^2 \phi(\bold{r}, t)
    \label{diffusion}
\end{equation}
\\*The above equation differs from the form of Tiribocchi et al. \cite{ref18} in that we have omitted the Ginzburg-Landau-like terms which they needed to study phase separation. In addition, we have neglected a non-linear term in the concentration current that leads to a term in Eq. \eqref{diffusion} that is of higher order in the gradients than the convective non-linearity in that equation. Effectively in each equation we have retained the leading non-linearity in a gradient expansion. This will allow us to obtain a generalized Lorenz model at the end of this section.

The setup we envisage consists of two square parallel plates of side $L_{\perp}$, separated by a distance $L$ and placed parallel to the $X-Y$ plane. We work in the limit of $ L_{\perp} \gg L$ (infinite aspect ratio).

The active matter gradient is maintained at a constant value $\frac{\Delta \phi_0}{L}$ across the plates. In this non equilibrium steady state (NESS), we have 
\begin{subequations}
\begin{equation}
    \phi_0(\bold{r}) = \phi_{00} +  z \frac{\Delta \phi_0}{L} \label{avgrad1}
\end{equation}
and, 
\begin{equation}
    \bold{u}_0(\bold{r}) = 0 \label{avgrad2}
\end{equation}
\end{subequations}
\\*In this steady state there is a constant particle current $\bold{j} = -D \bold{\nabla} \phi = - D \frac{\Delta \phi_0}{L} \bold{\hat{z}}$ across the system. For the motionless state of the fluid, we require the pressure gradient to vanish.

The first question that we ask is whether this NESS is stable against small perturbations. If $\delta u_i(\bold{r}, t)$ and $\delta \phi(\bold{r}, t)$ are the perturbations around $u_0(\bold{r})$ and $\phi_0(\bold{r})$, then to linear order in $\delta u_i$ and $\delta \phi$, we have 

\begin{equation}
    \partial_t \delta u_i  - \nu \nabla^2 \delta u_i = - \partial_i \delta p - \zeta \frac{\Delta \phi_0}{L} \left (\frac23 \partial_z \partial_i \delta \phi + \nabla^2 \delta \phi \right )
    \label{pert1}
\end{equation}
\\*Imposing the incompressibility constraint $\partial_i \delta u_i = 0$ from Eq. \eqref{goveq1} leads to 

\begin{equation}
    \nabla^2 \delta p = - \frac53 \zeta \frac{\Delta \phi_0}{L} \partial_z \nabla^2 \delta \phi
    \label{pert2}
\end{equation}
\\*Taking the Laplacian of Eq. \eqref{pert1} and using Eq. \eqref{pert2} leads to 

\begin{equation}
    \nabla^2 \left (\partial_t \delta u_i - \nu \nabla^2 \delta u_i \right ) =  \zeta \frac{\Delta \phi_0}{L} \left(\partial_z \partial_i \nabla^2 \delta \phi - \nabla^4 \delta \phi \delta_{i^3} \right) 
    \label{pert3}
\end{equation}
\\*The concentration equation yields 

\begin{equation}
    (\partial_t - D \nabla^2)\delta \phi = - \frac{\Delta \phi_0}{L} \delta w
    \label{conceq}
\end{equation} where $\delta w$ is the z-component of $\boldsymbol{\delta}\bold{ u}$. Consequently, in Eq. \eqref{pert3} we use only the z-component which can be written as 

\begin{equation}
    (\partial_t \delta w - \nu \nabla^2 \delta w) = - \zeta \frac{\Delta \phi_0}{L} \nabla^2_{\perp} \delta \phi
    \label{pert4}
\end{equation} where $\nabla^2_{\perp} = \nabla^2 - \frac{\partial^2}{\partial z^2}$. Eliminating $\delta \phi$ between Eqs. \eqref{conceq} and \eqref{pert4} we get

\begin{equation}
    (\partial_t - D \nabla^2)(\partial_t - \nu \nabla^2)\delta w = \zeta \left( \frac{\Delta \phi_0}{L} \right )^2 \nabla^2_{\perp} \delta w
    \label{weqn}
\end{equation}

The vertical velocity and its second derivative are taken to vanish at the two plates (stress-free boundary conditions) and because of the infinite extent in the $X-Y$ plane, we consider a periodic solution in the horizontal plane. The solution will have the form 

\begin{equation}
    \delta w(\bold{r}, t) = \sum_{n} a_n e^{ik_{\perp} \cdot r_{\perp}} \sin \left ( \frac{n \pi z}{L} \right ) e^{\lambda_n (k_{\perp}) t}
    \label{wsoln}
\end{equation}
and inserting in Eq. \eqref{weqn}, we have for each $n$,

\begin{equation}
\begin{aligned}
    \left [ \lambda_n + D \left ( \frac{n^2 \pi^2}{L^2} + k^2_{\perp} \right ) \right ] \left [ \lambda_n + \nu \left ( \frac{n^2 \pi^2}{L^2} + k^2_{\perp} \right ) \right ] \\
    = - \zeta \left( \frac{\Delta \phi_0}{L} \right )^2 k^2_{\perp} 
    \end{aligned}
    \label{lambdan}
\end{equation}
Instability would occur if for a given $n$ and $k_{\perp}$ it is possible for $\lambda_n(k_{\perp})$ to become zero or positive. It is immediately clear that for $\lambda_n$ to be zero or positive, we need $\zeta <0 $. Hence for large negative values of the activity coefficient, we can have the state given by Eqs. \eqref{avgrad1} and \eqref{avgrad2} become unstable. Writing $-\zeta$ as $\vert \zeta \vert$, we find the condition of $\lambda_n$ becoming zero as 

\begin{equation}
    \vert \zeta \vert \left( \frac{\Delta \phi_0}{L} \right )^2 = D \nu \left( \frac{\frac{n^2 \pi^2}{L^2}  + k^2_{\perp}}{k^2_{\perp}} \right )^2
    \label{cond1}
\end{equation}
Clearly, the lowest value of $n$ for which the instability can occur is $n=1$ and the critical value of $\vert \zeta \vert$ is 

\begin{equation}
    \vert \zeta \vert \left( \frac{\Delta \phi_0}{L} \right )^2  \frac{L^2}{D\nu} =    \frac{( \pi^2  + k^2_{\perp}L^2)^2}{L^2 k^2_{\perp}} 
    \label{criticalzeta}    
\end{equation}
We define $N\equiv \vert \zeta \vert  \frac{(\Delta \phi_0)^2}{D\nu}$ as the analogue of the Rayleigh number in fluids. This activity Rayleigh number $N$ has a minimum value when $k_{\perp} L = \pi$ and hence the critical value $N_c$ for the "convective instability" is 
\begin{equation}
    N_c = 4\pi^2
    \label{critN}
\end{equation}
We have, thus, arrived at a situation which is completely analogous to hydrodynamic convection. The pure conduction state when a fluid is subjected to a constant temperature gradient loses stability above a critical Rayleigh number (the critical value is $27\frac{\pi^4}{4}$ for the boundary conditions that we have used) and a steady convective state becomes stable. The Rayleigh number is the dimensionless variable $Ra = \frac{\alpha g}{D \nu} \left( \frac{\Delta T}{L} \right ) L^4$ with the $D$ standing for heat diffusivity (since the advected variable is heat), $\alpha$ is the expansion coefficient and $g$ is the acceleration due to gravity. The analogue of $\Delta \phi_0$ is $\Delta T$ and the analogue of $\vert \zeta \vert$ is $\alpha g$. 

It should be noted that for physical reasons, only positive $\Delta T$ (heated from below) can cause convective instability while in the present scenario, any sign of the gradient will lead to an instability if $N$ exceeds the critical value. 

In fluids, as the Rayleigh number is increased beyond a critical value, the steady convection becomes oscillatory as the roll-structure begins to undulate. With further increase of Rayleigh number, the oscillatory state makes a transition to a more complicated time dependence and eventually to a soft turbulence (chaos) followed by a hard turbulence characterized by definite scaling behavior. 
To see this transition in a tractable model, Lorenz carried out a Galerkin truncation of the hydrodynamic equations choosing the three most relevant modes. The resulting Lorenz model---a set of three ordinary non-linear differential equations has become one of the most studied examples in dynamical system theory for complex behavior. Motivated by Lorenz's work, we construct a similar dynamical system model for this instability. 

The simplest set of variables that we can take is those corresponding to a set of cylindrical rolls with the axis of the cylinder in the $y$-direction. This makes all the fields $\bold{u}(\bold{r}, t)$ and $\phi(\bold{r}, t)$ independent of the $y$-coordinate. The choice 

\begin{subequations}
\begin{equation}
        u_z = A(t)\cos{\pi x}\sin{\pi z} 
        \label{ansatza}
\end{equation}
\begin{equation}
    u_x = -A(t) \sin{\pi x} \cos{\pi z} 
    \label{ansatzb}
\end{equation}
\begin{equation}
    \delta \phi = \phi - \phi_0(\bold{r}) = B(t)\cos{\pi x}\sin{\pi z} + C(t) \sin{2\pi z} 
    \label{ansatzc}
\end{equation}
\end{subequations}
is the minimal nontrivial model. As shown in Ref. \cite{ref20}, this choice leads to the Lorenz model for driven active matter in the form

\begin{subequations}
    \begin{equation}
        \dot{X} = \sigma(-X+rY+rYZ) \label{system1}
    \end{equation}
    \begin{equation}
        \dot{Y} = -XZ + X - Y  \label{system2}
    \end{equation}
    \begin{equation}
        \dot{Z} = -2Z + XY  \label{system3}
    \end{equation}
\end{subequations}
\\*In the above, $\sigma = \frac{\nu}{D}$ is the inverse Schmidt number and $r = \frac{N}{N_c}$ is the control parameter. It should be noted that the inverse Schmidt number in this model will be much greater than the Prandtl number in the usual Lorenz model since the particle diffusion coefficient $D$ is generally much smaller than the heat diffusion coefficient. The difference with the standard Lorenz model lies in the non-linear term in Eq. \eqref{system1}. This term is the active matter contribution. 

Using an ansatz, we arrive at the following Lyapunov function for the Lorenz model written down above

\begin{equation}
    V(X, Y, Z) = X^2 + 2\sigma r Y^2 + \sigma r (Z-3)^2
    \label{Lyapfun}
\end{equation}
with $$\dot{V} = 2\left [ - \sigma X^2 - 2 \sigma r Y^2 - 2\sigma r (Z - \frac32)^2 + \frac92 \sigma r \right ]$$ 

Clearly, $\dot{V} < 0$ outside the ellipsoid $\sigma X^2 + 2 \sigma r Y^2 + 2 \sigma r (Z-\frac32)^2 = \frac92 \sigma r$ and hence all trajectories of the active matter Lorenz model remain bounded for all time. In the next section we will study the fixed points of this Lorenz model and their stability.

\section{Fixed Points and Bifurcations}
\label{sec3}

The fixed points of Eqs. \eqref{system1} --- \eqref{system3} are  \begin{enumerate}
    \item The trivial fixed point \begin{equation}
    X=Y=Z=0  \label{fixed1}
\end{equation}
    \item Two non-trivial fixed points at \begin{subequations}
\label{fixed2}
\begin{equation}
    X_0 = \pm \sqrt{2}\left [ (r-1) \pm \sqrt{(r-1)^2+(r-1)} \right ]^{\frac12}
    \label{fixed2a}
\end{equation}
\begin{equation}
    Y_0 = \frac{X_0}{1+\frac{X_0^2}{2}}
    \label{fixed2b}
\end{equation}
\begin{equation}
    Z_0 = \frac{X_0 Y_0}{2}
    \label{fixed2c}
\end{equation}
\end{subequations}
\end{enumerate}
The fixed point \eqref{fixed1} corresponds to the state where there is no flow and a static concentration gradient exists between the plates. Linearizing about this fixed point, we have the perturbations $\delta X, \delta Y, \delta Z$ follow the dynamics 

\begin{subequations}
    \begin{equation}
        \dot{\delta X} = \sigma (-\delta X  + r \delta Y)
        \label{trivialperta}
    \end{equation}
    \begin{equation}
        \dot{\delta Y} = \delta X - \delta Y
        \label{trivialpertb}
    \end{equation}
    \begin{equation}
        \dot{\delta Z} = -2 \delta Z
        \label{trivialpertc}
    \end{equation}
\end{subequations}
\\*One of the eigenvalues (growth rate of perturbation) is clearly $-2$ and the other two follow from 

\begin{equation}
    \begin{vmatrix}
        \lambda + \sigma & - \sigma r \\
        -1 & \lambda + 1 \\
    \end{vmatrix} = 0
    \label{dettriv}
\end{equation}
which yields 

\begin{equation}
    2 \lambda = - (1 + \sigma) \pm \sqrt{(1+\sigma)^2 + 4 \sigma (r-1)}
    \label{lambdatriv}
\end{equation}
\\*Clearly, this fixed point is stable for $r<1$ and unstable for $r>1$. The zero-velocity state consequently becomes unstable for $r>1$ and one has the steady convection roll state given by the fixed point \eqref{fixed2}. The values of $X_0$ and $Y_0$ correspond to the amplitudes of the roll-like solution while $Z_0$ is the Nusselt number for the problem. It describes the constant rate at which the active matter is transported from one plate to another. 

Unlike the Rayleigh-Bernard case, the convective transport here is proportional to $(r-1)^{\frac12}$ for $r>1$ which is different from the $r-1$ dependence of the usual Lorenz model. Similarly, for $r \gg 1$, the transport saturates in this case as opposed to the continuous increase in the standard situation.

We now look at the question of stability of fixed point \eqref{fixed2}. Writing $X = X_0 + \delta X$, $Y = Y_0 + \delta Y$ and $Z = Z_0 + \delta Z$ and linearizing in $\delta X, \delta Y$ and $\delta Z$, we arrive at the following cubic for the eigenvalue $\lambda$

\begin{equation}
\begin{aligned}
    \lambda^3 + \lambda^2 (3 + \sigma) + \lambda \left (2 + 3 \sigma + X_0^2 - \sigma r (1- Z_0^2) - \sigma r Y_0^2 \right ) \\
     + 2 \sigma + \sigma X_0^2 + 2 \sigma r (Z_0^2 - 1) + 4 \sigma r Z_0^2 - \sigma r Y_0^2 = 0
    \end{aligned}
    \label{cubic1}
\end{equation}

For $r = 1 + \epsilon$ ($\epsilon \ll 1$ but positive) it is clear that all the three roots are real and negative. As one increases $r$ beyond the critical value of $1$, the stable fixed points change from nodes to spirals and eventually become unstable via a Hopf bifurcation at $r = r_H$. At the Hopf point, one root is negative and the other two roots become pure imaginary, i.e. $\pm i \omega_H$.

\begin{figure}[t]
    \centering
    \includegraphics[width = 72 mm]{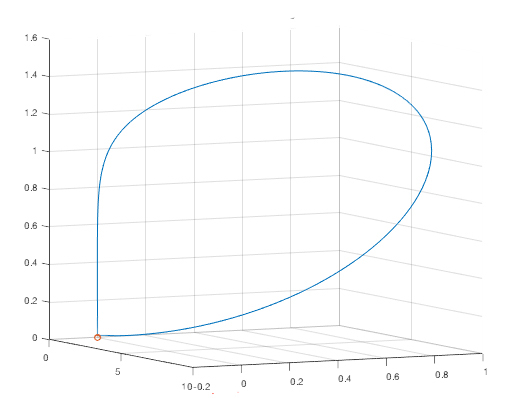}
    \caption{Homoclinic orbit at $\sigma = 10$ and $r = 5.45$. The orbit takes off from the origin along the unstable eigendirection in the X-Y plane and returns to it along a stable one (Z-axis).}
    \label{fig:homo}
\end{figure}

\begin{figure}[t]
    \centering
    \includegraphics[width = 90 mm]{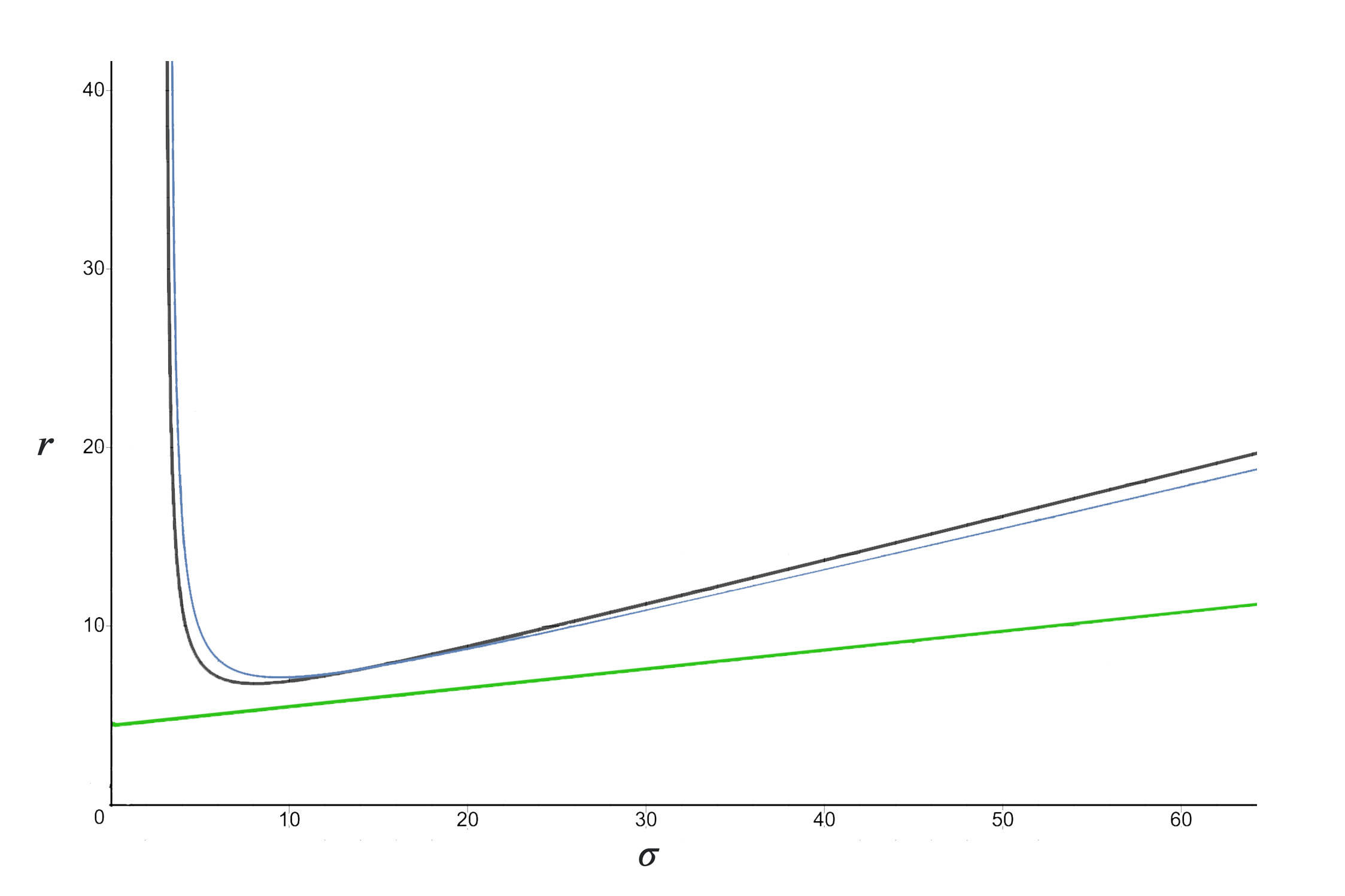}
    \caption{The blue line represents the numerical solution for $r_H$ , black line represents the approximated $r_H$ and green line represents $r_0$, all as functions of $\sigma$}
    \label{fig:critical r}
\end{figure}

Before discussing what happens at $r = r_H$, we need to ask if a homoclinic bifurcation occurs for $1 < r < r_H$, as in the standard Lorenz model \cite{ref23}---\cite{ref25}. To find the homoclinic bifurcation we recall that the origin is an unstable fixed point (for $r>1$) with a one-dimensional unstable manifold and a two-dimensional stable manifold (the eigenvalues are $\lambda_{1, 2}$ from \eqref{lambdatriv} and $\lambda_3 = -2$). At the homoclinic point $r_0$, an orbit orbit leaves the origin along the unstable eigendirection, loops around the nearest stable fixed point and returns to the origin along the stable manifold. The homoclinic point $r_0$ as a function of $\sigma$ has been determined numerically (a typical orbit at $r=r_0 = 5.45$ and $\sigma = 10$ is shown in Fig. \ref{fig:homo}) and have been plotted as the green curve in Fig. \ref{fig:critical r}. The interesting thing to note is that $r_0 < r_H$ for all values of $\sigma$. In fact for large $\sigma$, both $r_H$ and $r_0$ grow almost linearly with $\sigma$, with $r_0$ having a smaller slope as can be seen in Fig. \ref{fig:critical r}. This means that a homoclinic bifurcation always occurs below the Hopf point and there could exist an unstable limit cycle in the region $r_0< r < r_H$, that is created from the homoclinic bifurcation at $r_0$ and vanishes at the Hopf bifurcation at $r_H$ (thus making it a sub-critical one). However this is not necessarily true for all $\sigma$, as we shall show soon.

Returning to the Hopf bifurcation point at $r = r_H$, we have 

\begin{subequations}
    \begin{equation}
        \omega_H^2 = 2 + X_0^2 + 3 \sigma - \sigma r_H (1-Z_0^2) - \sigma r_H Y_0^2 
        \label{cubic2a}
    \end{equation}
    \begin{equation}
        (3 + \sigma) \omega_H^2 = \sigma (2 + X_0^2) + 2 \sigma r_H(3 Z_0^2 - 1) - \sigma r_H Y_0^2
        \label{cubic2b}
    \end{equation}
\end{subequations}
We need to find $r_H$ and $\omega_H$ as functions of $\sigma$, using the $X_0, Y_0, Z_0$ given in Eqs. \eqref{fixed2a}---\eqref{fixed2c}. This is difficult and so we use a large-$\sigma$ approximation since we expect $\sigma$ to be large for this problem. It is easy to check that for $\sigma \gg 1$, $r_H \approx \frac{\sigma}{4}$ and hence $r_H$ is large which in turn allows us to use asymptotic forms for $X_0, Y_0, Z_0$ which we write as follows (using $\epsilon = r - 1$ as the natural variable)

\begin{equation}
\label{asymp}
    \begin{gathered}
        X_0^2 = 4 \epsilon + 1 + O \left (\frac{1}{\epsilon}\right ) \\
        Y_0^2 = \frac{1}{\epsilon} \left [1 - \frac{5}{4\epsilon} + O\left(\frac{1}{\epsilon^2} \right ) \right ] \\
        Z_0^2 = 1 - \frac{1}{\epsilon} + \frac{1}{\epsilon^2} + O\left(\frac{1}{\epsilon^3} \right )
    \end{gathered}
\end{equation}
Using the above forms of $X_0, Y_0$ and $Z_0$ in Eqs. \eqref{cubic2a} and \eqref{cubic2b}, we obtain for $\sigma \gg 1$

\begin{equation}
    r_H = 1 + \frac{\sigma^2 + 6\sigma + \sigma\sqrt{\sigma^2 + 16\sigma + 24}}{8(\sigma - 3)} \label{critical r}
\end{equation}
and

\begin{equation}
    \omega_H^2 = 8\sigma (r_H-1) \label{critical w}
\end{equation}

The accuracy of the approximate $r_H$ of Eq. \eqref{critical r} can be assessed from Fig. \ref{fig:critical r} where we show the values of $r_H$ obtained from Eq. \eqref{critical r} and from a numerical solution of Eqs. \eqref{cubic2a} and \eqref{cubic2b} as a function of the inverse Schmidt number $\sigma$. It should be noted that for all $\sigma$, $r_0 < r_H$.

The emergence of a limit cycle with frequency $\omega_H$ at $r = r_H$ implies that at $ r = r_H$, the asymptotic solution for $X, Y, Z$ must have the form $(X, Y, Z) = (A, B, C) e^{\pm i \omega_H t}$ where $A, B$ and $C$ are constants. The fact that we have an instability at $r = r_H$ means that if we consider $r = r_H + \Delta r$ with $0 < \Delta r \ll r_H$, then the amplitudes $A, B, C$ will evolve in time as $e^{\lambda t}$ with $\lambda > 0$. We establish this behavior in Appendix A. To end this section, we check whether the growth saturates with increasing time (forward Hopf bifurcation). Mclaughlin and Martin \cite{ref26} had used a solvability based approach for the original Lorenz model which did show that the bifurcation was backward for $\sigma = 10$, but missed the fact that it could be forward for very high $\sigma$. This change with $\sigma$ was captured by the renormalization group treatment of Das et al. \cite{ref27}. In this work, we show that a Krylov-Bogolyubov approach (generally reserved for two-dimensional dynamical systems) is the most direct route to the amplitude equation, the details of which are provided in Appendix A. Here we simply quote the final result \eqref{finalamp} for the slow dynamics of the amplitude $A(t)$ of the limit cycle about the fixed point $(X_0, Y_0, Z_0)$ as

\begin{equation}
    \label{final}
    \left [ 4 \sigma - 2 i \omega_H \sigma  \right ] \frac{dA}{dt} = 8 \sigma (\Delta r) A - 4 A^3 \sigma + O \left (\frac1{\sigma} \right )
\end{equation}

\begin{figure*}[ht]
    \centering
    \includegraphics[width = 120 mm]{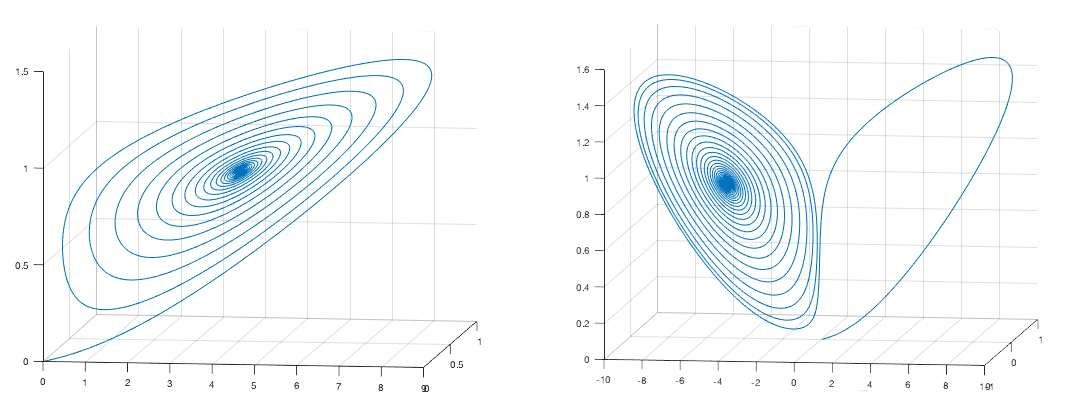}
    \caption{(Left) Trajectories starting along the unstable manifold land on the nearest non-trivial fixed point for $\sigma = 10$, $r = 5$. (Right) Trajectories starting along the unstable manifold wind around the nearest non-trivial fixed point and land on the other one for $\sigma = 10$, $r = 6$. From this it is clear that at $\sigma= 10\text{, } 5 < r_0 < 6$.}
    \label{fig:homobif}
\end{figure*}

The bifurcation is forward (supercritical) for $\sigma \gg 1$ and Eq. \eqref{final} shows that for $0< \Delta r \ll r_H$, the radius of the limit cycle will grow as $\sqrt{\Delta r}$ independent of $\sigma$. The correction term in Eq. \eqref{final} has a positive sign which indicates that below a certain critical $\sigma$, the Hopf bifurcation will be backward (subcritical). The case $\sigma = 10$ reported in Ref. \cite{ref20} is consistent with a backward Hopf bifurcation. There, a limit cycle was observed for $r < r_H$ provided the initial conditions were sufficiently close to the center of the limit cycle. We note that transitions between different dynamical states have been studied recently in a variety of active matter systems, like fluid flocks with inertia \cite{ref28}, droplet growth in scalar active matter \cite{ref29} and hydrodynamics of active defects \cite{ref30}. We now present the detailed numerical results in Sec. IV.

\section{Numerical Results and Discussion}
\label{sec4}

\begin{figure}
    \centering
    \includegraphics[width = 80mm]{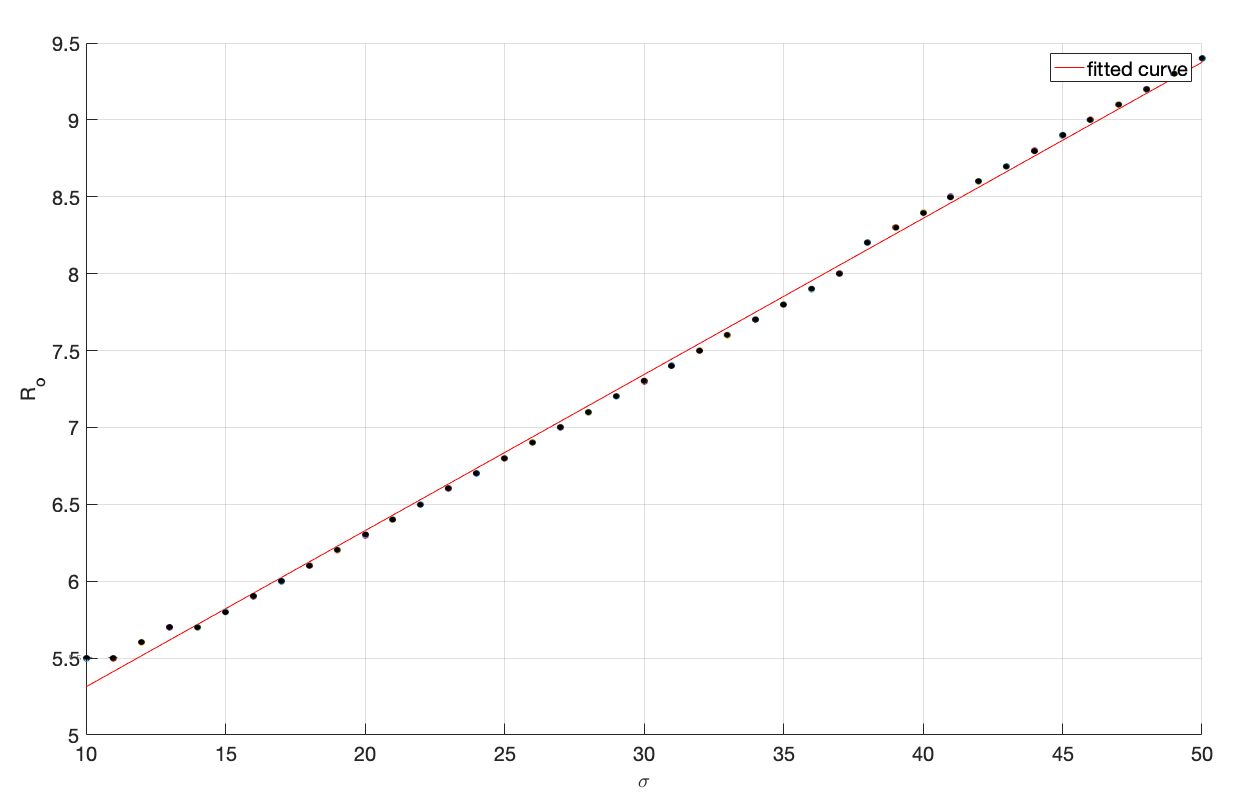}
    \caption{The critical point for the homoclinic bifurcation, $r_0$, as a function of $\sigma$}
    \label{fig:homonum}
\end{figure}

Similar to the Lorenz model, a homoclinic bifurcation occurs at $r = r_0$ in our system. The bifurcation can be identified by the transition shown in Fig. \ref{fig:homobif}. At the critical value $r = r_0$ there is a homoclinic orbit as shown in Fig. \ref{fig:homo} where the trajectory comes back to the origin. Using this and elementary iterative techniques, we numerically calculate $r_0$ as a function of $\sigma$. The plot is shown in Fig. \ref{fig:homonum}. As can be seen, for large $\sigma$, $r_0$ grows almost linearly. Plotting this curve in Fig. \ref{fig:critical r}, we see that $r_0 < r_H$ for all $\sigma$. Hence we always have a homoclinic bifurcation below the Hopf bifurcation.

\begin{figure}
    \centering
    \includegraphics[width = 65 mm]{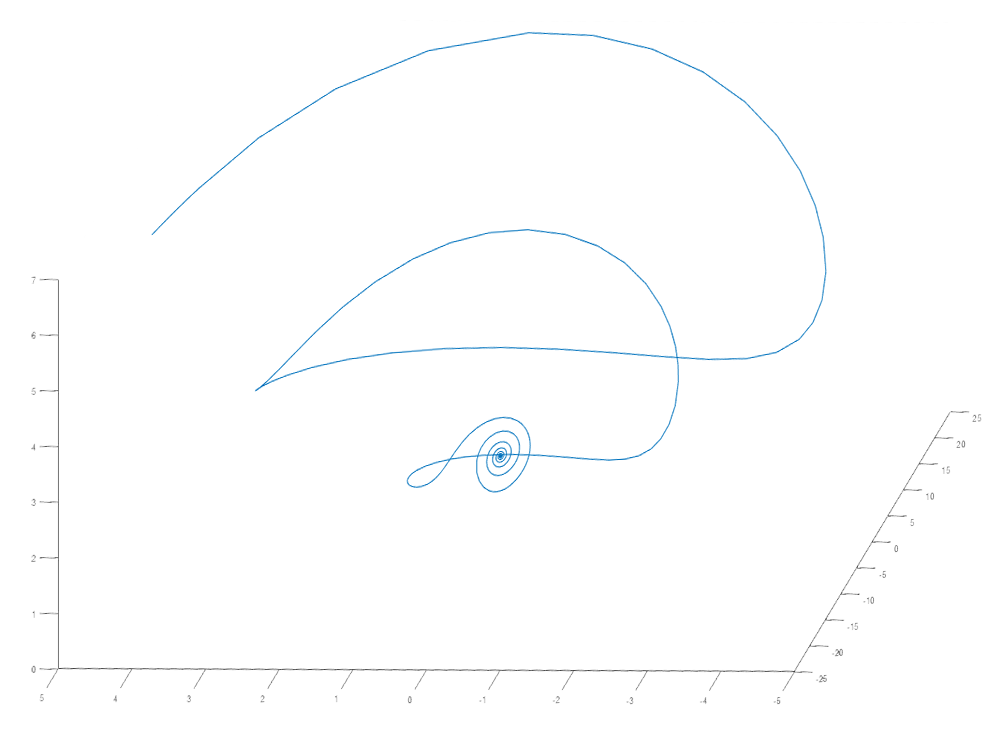}
    \caption{Stable spirals converging to fixed point at $\sigma = 10$ and $r < r_H$} 
    \label{fig:stablespiral2}
\end{figure}

\begin{figure}
    \centering
    \includegraphics[width = 70mm]{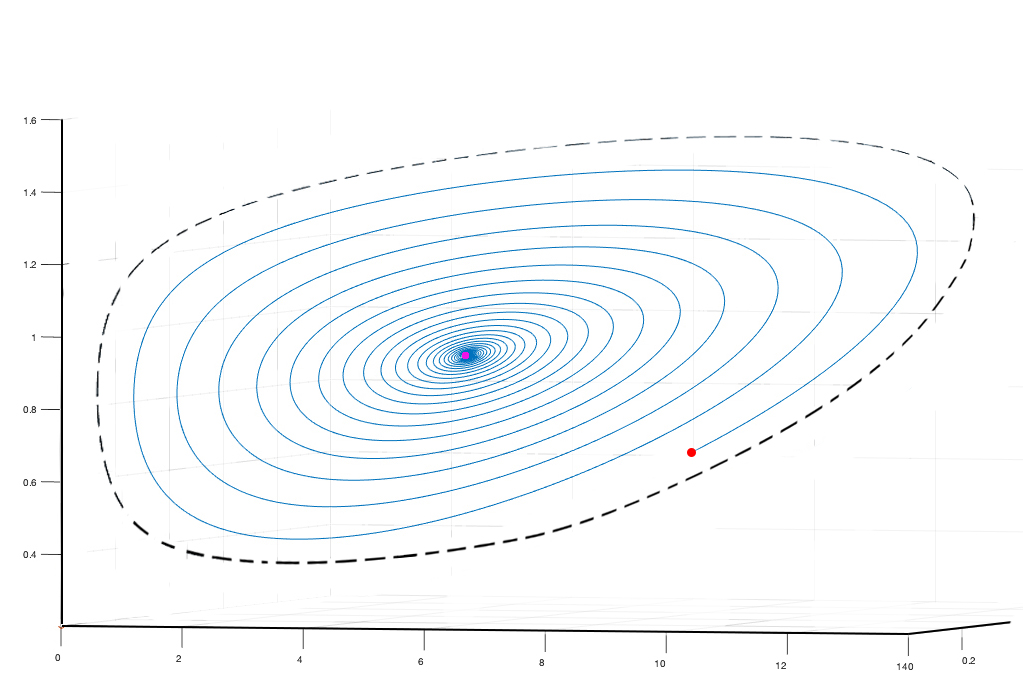}
    \caption{Unstable limit cycle (black) for $\sigma = 10$ and $r = 6$. The other trajectory (blue) starting at the red point spirals to the stable fixed point (purple)}
    \label{fig:unlc2}
\end{figure}

\begin{figure}
    \includegraphics[width = 70mm]{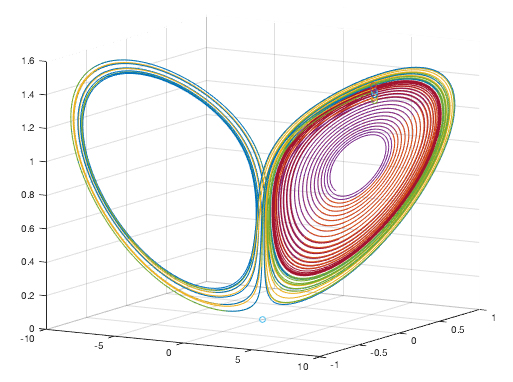}
    \caption{For low $\sigma$ and $r_0 < r < r_H$, initial conditions that are very close to the unstable limit cycle either spiral to the stable fixed point (red) or settle on the strange attractor (green, blue and yellow). The unstable limit cycle exists at the boundary of the red and green trajectories. }
    \label{fig:smol2}
\end{figure}

\begin{figure}
    \centering
    \includegraphics[width = 60 mm]{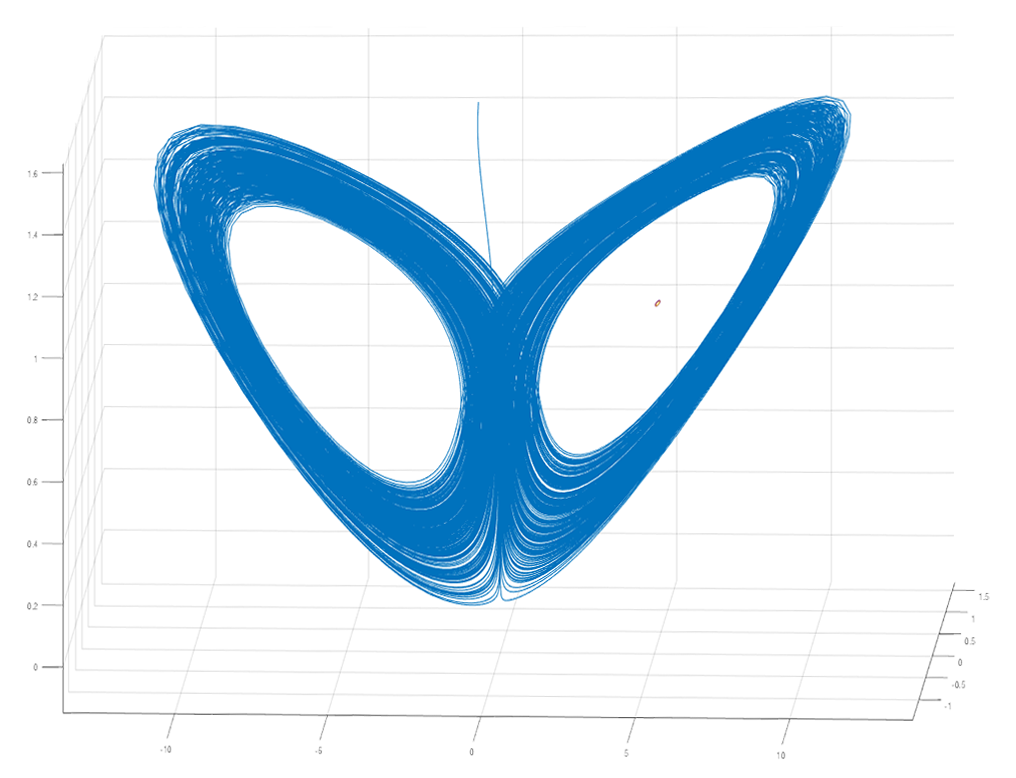}
    \caption{Lorenz-like strange attractor at $r = 10$ and $\sigma = 10$. The red dot in the right half is one of the (now) unstable fixed points.} 
    \label{fig:attractor}
\end{figure}

\begin{figure}
    \includegraphics[width = 70mm]{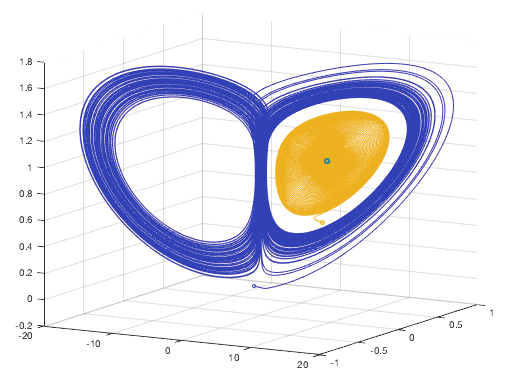}
    \caption{At $\sigma = 30$ (high sigma), $r = 10$ ($r_0 < r < r_H$) there coexist a chaotic strange attractor and non-trivial stable fixed points. Initial conditions close to the fixed points spiral into them (yellow), whereas others end up on the strange attractor (blue).}
    \label{fig:smol13}
\end{figure}

For low values of $\sigma$, our system is expected to behave like the standard Lorenz model. For $r < 1$ the origin is globally stable and the only fixed point. At $r = 1$ a pitchfork bifurcation occurs and for $1< r < r_H$ the two non-trivial fixed points given by Eq. \eqref{fixed2} are locally stable and trajectories starting sufficiently close to them spiral down to them as in Fig. \ref{fig:stablespiral2}. However the homoclinic bifurcation at $r_0$ leads to the formation of an unstable limit cycle as shown for $\sigma = 10$ in Fig. \ref{fig:unlc2} that exists for $r_0 < r < r_H$. Integrating numerically in this region, (using a Runge-Kutta $(4,5)$ scheme with an absolute and relative error tolerance of $10^{-6}$) we get chaotic trajectories coexisting with the spirals and the unstable limit cycle as shown in Fig. \ref{fig:smol2}. Since the limit cycle is unstable, even initial conditions very near it either settle on the fixed points or on the chaotic attractor. The limit cycle grows smaller on increasing $r$ from $r_0$ and vanish at $r = r_H$. For $r > r_H$, there are no periodic orbits or stable fixed points in the phase space. All trajectories end up on the Lorenz-like strange attractor shown in Fig. \ref{fig:attractor}. The same behaviour is shown for $\sigma = 12, 15, 20,$ etc.

\begin{figure*}[]
\subfloat[r = $12.1$ \label{fig:trans1}]{ \includegraphics[width = 40mm]{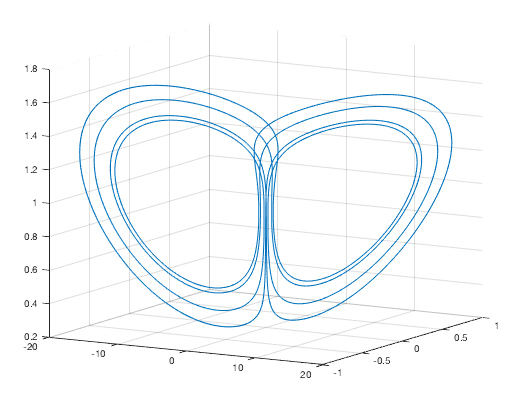}}
\subfloat[r = $12.4$ \label{fig:trans2}]{ \includegraphics[width = 40mm]{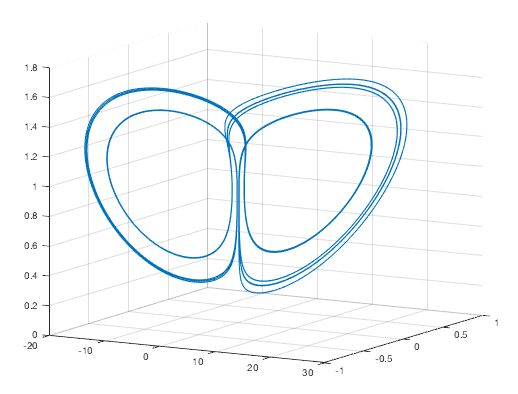}}
\subfloat[r = $12.5$ \label{fig:trans3}]{ \includegraphics[width = 40mm]{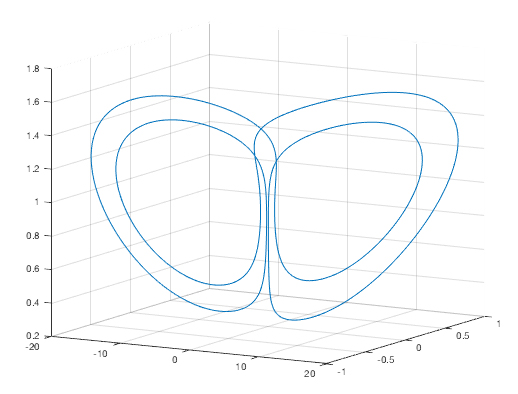}}
\subfloat[r = $15$ \label{fig:trans4}]{ \includegraphics[width = 40mm]{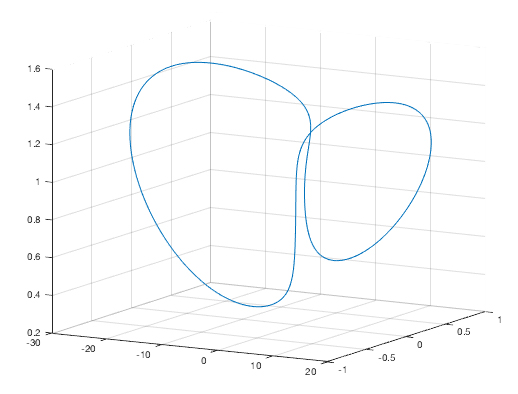}}

\subfloat[r = $16$ \label{fig:trans5}]{ \includegraphics[width = 40mm]{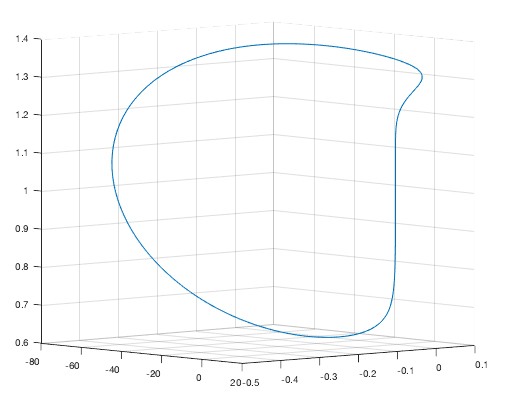}}
\subfloat[r = $19$ \label{fig:trans6}]{ \includegraphics[width = 40mm]{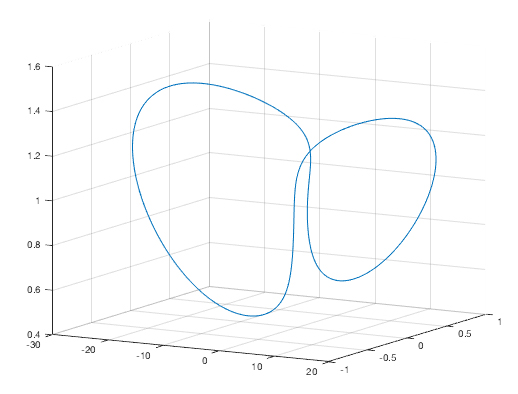}}
\subfloat[r = $38$ \label{fig:trans7}]{ \includegraphics[width = 40mm]{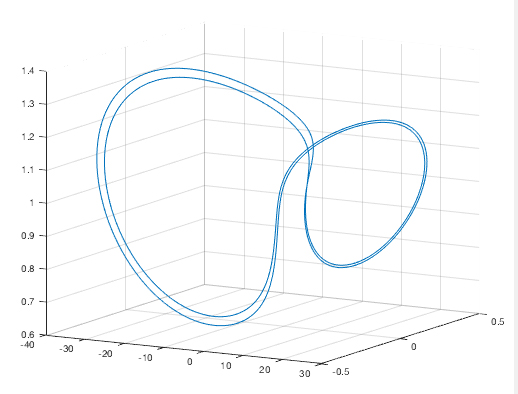}}
\subfloat[r = $41$ \label{fig:trans8}]{ \includegraphics[width = 40mm]{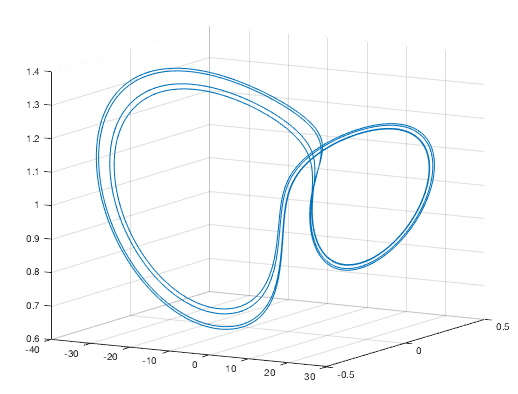}}

\subfloat[r = $43$ \label{fig:trans9}]{ \includegraphics[width = 40mm]{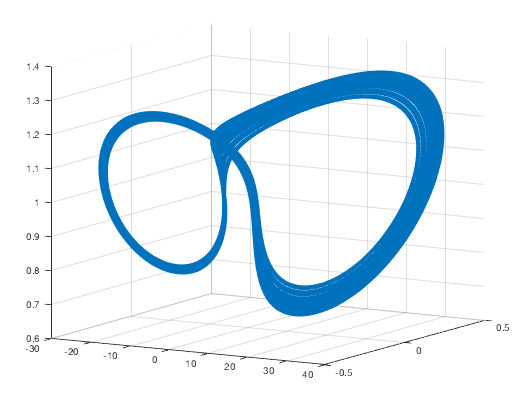}}
\subfloat[r = $45$ \label{fig:trans10}]{ \includegraphics[width = 40mm]{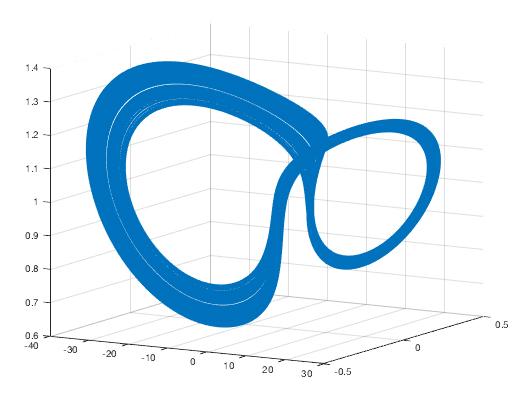}}
\subfloat[r = $50$ \label{fig:trans11}]{ \includegraphics[width = 40mm]{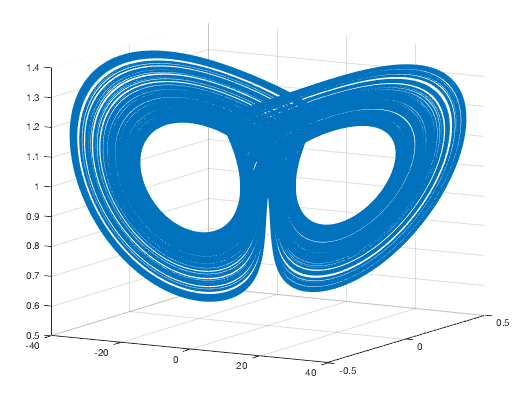}}
\subfloat[r = $60$ \label{fig:trans12}]{ \includegraphics[width = 40mm]{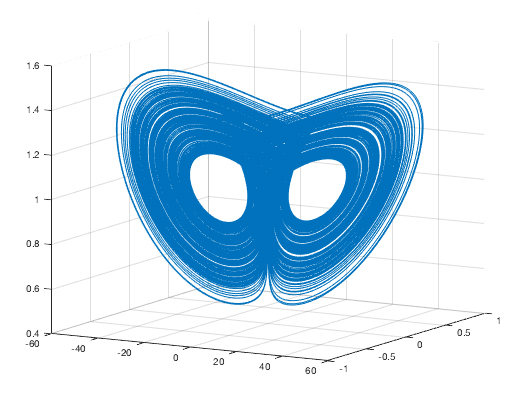}}
\caption{\protect\subref{fig:trans1} Stable high-period ($64 \frac{2 \pi}{\omega_H}$) limit cycle just above $r_H$, \protect\subref{fig:trans2} Period $8\frac{2 \pi}{\omega_H}$ limit cycle at $r = 12.4$, \protect\subref{fig:trans3} Period $4\frac{2 \pi}{\omega_H}$ limit cycle at $r = 12.5$,  \protect\subref{fig:trans4} Period $2\frac{2 \pi}{\omega_H}$ limit cycle at $r = 15$,  \protect\subref{fig:trans5} Period $\frac{2 \pi}{\omega_H}$ limit cycle at $r = 16$,  \protect\subref{fig:trans6} Period $2\frac{2 \pi}{\omega_H}$ limit cycle at $r = 19$,   \protect\subref{fig:trans7} --- \protect\subref{fig:trans10} Limit cycle continues to period double and becomes the strange attractor in \protect\subref{fig:trans11} and \protect\subref{fig:trans12}.}
\label{fig:smallsec3}
\end{figure*}

For higher $\sigma$ the behaviour of the system is starkly different. From the theoretical work in Sec. III we expect the Hopf bifurcation to be forward. For $r < r_H$, the non-trivial fixed points given by Eq. \eqref{fixed2} are stable. Although a homoclinic bifurcation still occurs in the $r_0 < r < r_H$ region, there is no periodic orbit, unlike the low $\sigma$ case. Trajectories either spiral to the stable fixed point or end up on the Lorenz-like strange attractor as shown in Fig. \ref{fig:smol13}. At $r = r_H$ a stable high-period limit cycle is born whose transition to chaos is quite unique and is shown in detail for $\sigma = 30$ in Fig. \ref{fig:smallsec3}. The limit cycle, shown in Fig.  \ref{fig:trans1}, has a time period of $8.94$ which corresponds to $64\frac{2\pi}{\omega_H}$, where $\omega_H$ is supposed to be the Hopf frequency of Eq. \eqref{critical w}.  As $r$ is increased beyond $r_H$, the limit cycle undergoes a succession of period halving bifurcations, as can be seen in Fig. \ref{fig:trans2}, \ref{fig:trans3} and \ref{fig:trans4}, until we have a $\frac{2\pi}{\omega_H}$ period limit cycle at some $r$ as shown in Fig. \ref{fig:trans5}. As $r$ is further increased, this limit cycle starts to double its period as shown in Fig. \ref{fig:trans6}, \ref{fig:trans7}, \ref{fig:trans8}, \ref{fig:trans9}, \ref{fig:trans10} and ultimately becomes the Lorenz-like attractor as in Fig. \ref{fig:trans11} and \ref{fig:trans12}. Throughout this process, the limit cycle is globally stable and there exists no chaos in the region $r > r_H $ except after the Lorenz-like attractor is born. Beyond this point, the strange attractor becomes the only dominant feature of the system. This behaviour can be seen for $\sigma = 22, 23, 25, \dots , 30, 40, 50, \dots $ and even $\sigma = 80$

Interestingly, as $\sigma$ is increased the zone of the reverse bifurcations narrows and the transition of the $\frac{2 \pi}{\omega_H}$-period limit cycle into the strange attractor gets more delayed (occurs at even higher values of $r$) and in the $\sigma \to \infty$ limit, the system shows no chaos since \eqref{system1} gives $$\lim_{\sigma \to \infty} \frac{\dot{x}}{\sigma} = 0 = -X+rY+rYZ  \implies X = rY(1+Z)$$ and we are essentially left with a two-dimensional flow which can never be chaotic. 

The critical sigma to separate the low and high cases is  calculated numerically to be approximately $21$. For $\sigma > 21$ the high-sigma route is followed whereas for $\sigma < 21 $ the low $\sigma$ behaviour is observed. 

All these results are summed up in the illustrated plot in Fig. \ref{fig:finalplot}.

\section{Conclusion}
\label{sec5}

We have considered a setup where a definite gradient of active matter is maintained in a given direction across the fluid medium in which it is suspended. As the gradient is increased the fluid medium develops ``convective'' instabilities and exhibits an infinite sequence of bifurcations. We have used a low-order Galerkin truncation to capture the dynamics of the system exactly in a manner analogous to that employed by Lorenz \cite{ref4} for studying atmospheric flows. This new version of the Lorenz model, so obtained, has non-linear terms in all the three equations. We find that this Lorenz model can actually show one of the most desirable routes to turbulence---from a trivial steady-state to a non-trivial one, followed by a Hopf bifurcation to a periodic state and then a sequence of period doubling bifurcations until the flow loses all periodicity and shows the existence of a stable strange attractor. Very surprisingly, there is a small twist---the period doubling bifurcations leading to chaos as the control parameter is increased have a mirror image of bifurcations as the control parameter is decreased and this leads to the coexistence of a strange attractor and stable fixed points on the other side of the Hopf bifurcation point. For inverse Schmidt numbers below a critical value, the bifurcations are identical to what one sees in the traditional Lorenz model.

\appendix
\renewcommand{\theequation}{A.\arabic{equation}}
\section{Slow dynamics of limit cycle amplitude}
\label{Appen}
We begin by rewriting our Lorenz equations as a single differential equation. Using the notation $D \equiv \frac{d}{dt}$, we define variables $u, v, w$ as 

\begin{equation}
    u = X - X_0, \ \ v = Y - Y_0, \ \  w = Z - Z_0
\end{equation} and write Eqs. \eqref{system1}---\eqref{system3} as

\begin{subequations}
    \begin{equation}
        (D + \sigma) u = \sigma r (1+Z_0)v + \sigma r Y_0 w + \sigma r wv
        \label{syseqx}
    \end{equation}
    \begin{equation}
        (D+1)v = (1-Z_0)u - X_0 w - uw
    \end{equation}
    \begin{equation}
        (D+2)w = Y_0 u + X_0 v + uv
    \end{equation}
\end{subequations}

Combining into one equation, we have 

\begin{equation}
    \begin{aligned}
        Lu &=  \big{[}  (D + \sigma)(D+1)(D+2) + X_0^2 (D+\sigma) \\
        &- \sigma r (1-Z_0^2)(D+2) + 4 \sigma r Z_0^2 - \sigma r Y_0^2 (D+1) \big{]} u \\
         &= - \sigma r X_0(1+Z_0)uv - \sigma r (1+Z_0)(D+2)uw  \\
         &- 2Z_0\sigma r u w + \sigma r Y_0 (D+1)uv \\
         &+ \sigma r \left [(D+1)(D+2) + X_0^2\right] wv  
    \end{aligned}
    \label{combode1}
\end{equation}
We also note 
\begin{subequations}
    \begin{equation}
    \begin{aligned}
        \left [(D+1)(D+2) + X_0^2\right ] &v  \\
        = (1-Z_0)(D+2)u - 2Z_0 u &- X_0 u v - (D+2) uw
        \label{combode2}
        \end{aligned}
    \end{equation}
    and
    \begin{equation}
    \begin{aligned}    
        \left [(D+1)(D+2) + X_0^2\right ] &w  \\
        = Y_0(D+1)u + X_0(1-Z_0)&u - X_0 uw + (D+1)uv
        \label{combode3}
        \end{aligned}
    \end{equation}
\end{subequations}

We start exploring the region $r = r_H + \Delta r$ by ignoring the non-linear terms in Eq. \eqref{combode1}  and splitting the linear part as a part at $r = r_H$ and a part of $O(\Delta r)$. We get

\begin{equation}
\label{combode4}
    \begin{gathered}
        \Big [(D+ \sigma)(D + 1)(D+2) + X_{0c}^2 (D+\sigma) \\
        - \sigma r_H (1- Z_{0c}^2)(D+2) + 4 \sigma r_H Z_{0c}^2 - \sigma r_H Y_{0c}^2 (D+1) \Big ] u \\
         + \Bigg [ \frac{\partial X_{0c}^2}{\partial r} (D+ \sigma) - \sigma (1 - Z_{0c}^2)(D+2) + \sigma r_H \frac{\partial Z_{0c}^2}{\partial r}(D+2) \\
         + 4 \sigma Z_{0c}^2 + 4 \sigma r_H \frac{\partial Z_{0c}^2}{\partial r} - \sigma \frac{\partial}{\partial r }(r_H Y_{0c}^2) \Bigg ] \Delta r u = 0
    \end{gathered}
\end{equation}
Knowing that at $\Delta r = 0$, the solution is $Ae^{i \omega_H t}$ where $A$ is a constant, we write the solution for $\Delta r \ll r_H$ as $u = A(t)e^{i \omega_H t}$ where $A(t)$ is a slowly varying function of time which implies $\frac{dA}{dt}$ is $O(\Delta r)$. The first term on the left-hand side of Eq. \eqref{combode4} factors as $(D^2 + \omega_H^2)(D + 3 + \sigma)$ [see Eqs. \eqref{cubic1}, \eqref{cubic2a} and \eqref{cubic2b}] and in the second term the $D$ operator can be replaced by $+i\omega_H$. Thus we find,

\begin{equation}
    \begin{gathered}
        (D^2 + \omega_H^2)(D + 3 + \sigma) u + \Delta r \Bigg [(i\omega_H + \sigma) \frac{\partial X_{0c}^2}{\partial r} + \\
        \sigma r_H \frac{\partial Z_{0c}^2}{\partial r} (i\omega_H + 2) - \sigma (1- Z_{0c}^2)(i \omega_H + 2) + 4 \sigma Z_{0c}^2 \\
        + 4 \sigma r_H \frac{\partial Z_{0c}^2}{\partial r} - \sigma \frac{\partial r Y_{0c}^2}{\partial r}\Bigg ]u = 0
        \label{combode5}
    \end{gathered}
\end{equation}
In the first term, dropping the second and higher-order derivatives of $A(t)$, we have 
\begin{equation}
\label{combode6}
    \begin{aligned}
        &(D + 3 + \sigma)(D^2 + \omega_H^2)\left[A(t) e^{i \omega_H t}\right ] \\
        = &(D + 3 + \sigma)(DA)e^{i \omega_H t}\\
        = &2 \left[ -\omega_H^2 + i \omega_H (3 + \sigma)\right]\frac{dA}{dt} e^{i \omega_H t}
    \end{aligned}
\end{equation}

Now using the relations given in Eq. \eqref{asymp}, we approximate the second term of Eq. \eqref{combode5} as $$\Delta r \left [ 4 (i \omega_H + \sigma)  + 4 \sigma - i \sigma \omega_H\frac{1}{r_H} \right ] \approx 8 \Delta r  \sigma$$ to the leading order in $\sigma$ using $r_H \approx \frac{\sigma}{4}$. Thus, Eq. \eqref{combode5} becomes 

\begin{equation}
    2\left [-\omega_H^2 + i \omega_H (3 + \sigma)\right ] \frac{dA}{dt} + 8 \sigma \Delta r A = 0
    \label{combode7}
\end{equation}
which clearly shows that A increases exponentially with time for $\Delta r >0$ and hence there is an instability at $r = r_H$. 

To complete the story, we need to know the first non-linear contribution to Eq. \eqref{combode7} and determine if the growth saturates. To do this we need to go back to Eqs. \eqref{combode1} --- \eqref{combode3} and from the non-linear term on the right-hand side we extract the part with a time-dependence $e^{i\omega t}$. For this, we work with $u = A e^{i \omega t}$ ignoring the time-dependence of $A$ in taking the derivatives on the right hand side of Eqs. \eqref{combode1} --- \eqref{combode3}. This is allowed because the non-linear term itself will be $O(A^3)$ which will make $A^2$ of $O(\Delta r)$ if the growth saturates. We first find the lowest order $v$ and $w$  ($v_0$ and $w_0$) in terms of $u$ from the linear terms in Eqs. \eqref{combode2} and \eqref{combode3}. In the $\sigma \gg 1$ approximation, which keeps the algebra simple,

\begin{subequations}
    \begin{equation}
        v_0 = 2 A \frac{\sigma + 3 i \omega }{\sigma^2 + 9 \omega^2} e^{i \omega t} + O\left(\frac1{r_H}\right)
        \label{vzero}
    \end{equation}
    \begin{equation}
        w_0 = 2 \sigma^{\frac12} A \frac{4-i\omega}{\sigma^2 + 9\omega^2} e^{i \omega t} + O\left(\frac1{r_H}\right)
        \label{wzero}
    \end{equation}
\end{subequations}
The next step requires us to find the $O(A^2)$ terms which we denote by $u_1, v_1, w_1$ from Eqs. \eqref{combode2}, \eqref{combode3} and \eqref{syseqx}. Since the non-linearity is quadratic, the expected structure is

\begin{subequations}
    \begin{equation}
        u_1 = A^2 \left [ B_1 e^{2 i \omega t} + B_1^* e^{-2 i \omega t} + C_1\right ]
    \label{uone}    
    \end{equation}
    \begin{equation}
        v_1 = A^2 \left [ B_2 e^{2 i \omega t} + B_2^* e^{-2 i \omega t} + C_2\right ]
        \label{vone}
    \end{equation}
    \begin{equation}
        w_1 = A^2 \left [ B_3 e^{2 i \omega t} + B_3^* e^{-2 i \omega t} + C_3\right ]
        \label{wone}
    \end{equation}
\end{subequations}
The constants $B_2, B_3$ and $C_2, C_3$ can be found from the non-linear part of Eqs. \eqref{combode2} and \eqref{combode3} and $B_1, C_1$ from Eq. \eqref{syseqx}. To keep the algebra simple, we show the results in the $\sigma \gg 1$ limit only. The derivation will be outlined for $v_1$ and the results for $w_1$ and $u_1$ written down. For this purpose, we note the following large $\sigma$ approximations

\begin{equation}
    \label{approx}
    r_H \approx \frac{\sigma}{4} \text{, } \omega_H^2 \approx 2 \sigma \text{, } X_0^2 \approx \sigma \text{, } Y_0 \approx \frac{1}{\sqrt{r}} \text{, } Z_0 \approx 1 - \frac2{\sigma}
\end{equation}
Starting with Eq. \eqref{combode2}, we write 
\begin{equation}
\label{combode8}
    \begin{aligned}
         &[ (D + 1)(D + 2) + \sigma  ] v_1 \\
        = &- \sigma^{\frac12} u_0 v_0 - (D + 2)u_0 w_0 \\
        = &- \sigma^{\frac12} \left [ A B_1 e^{ 2 i \omega t} + 2 \operatorname{Re} (A B_1^*) \right ] \\
        &- (D+2) \left [ A C_1 e^{2 i \omega t} + 2 \operatorname{Re} (A C_1^*) \right ] \\
        = &- \sigma^{\frac12} A^2 \left [ 2 \frac{\sigma + 3 i \omega}{\sigma^2 + 9 \omega^2} e^{ 2 i \omega t} +  \frac{4 \sigma}{\sigma^2 + 9 \omega^2} \right ] \\
        &- \sigma^{\frac12} A^2 (D+2) \left [ 2 \frac{4 -  i \omega}{\sigma^2 + 9 \omega^2} e^{ 2 i \omega t} +  \frac{16 A^2}{\sigma^2 + 9 \omega^2} \right ] \\
    \end{aligned}
\end{equation}
We can now read off the coefficients $B_2$ and $C_2$ of Eq. \eqref{vone} from \eqref{combode8} as ($\sigma \gg 1$)

\begin{equation}
    B_2 \approx \frac{10}{7} A^2 \sigma^{-\frac32} \text{, } C_2 \approx -36 A^2 \sigma^{-\frac32}
    \label{B2C2}
\end{equation}
Similarly, starting with Eq. \eqref{combode3}, we arrive at 

\begin{equation}
    B_3 \approx \frac{282 A^2}{49 \sigma^2} \text{, } C_3 \approx -12 \frac{A^2}{\sigma^2}
    \label{B3C3}
\end{equation}
To find $u_1$ we turn to Eq. \eqref{syseqx} and write the second-order terms as 

\begin{equation}
\label{combode9}
    \begin{aligned}
        (D +  \sigma) u_1 &= \frac{\sigma^2}{2} v_1 + \frac{\sigma^{\frac32}}{2} w_1 + \frac{\sigma^2}{4} w_0 v_0 \\
        &= \frac{\sigma^2}{2} \left [ B_2 e^{ 2 i \omega t} + B_2^* e^{- 2 i \omega t}  + C_2 \right ] \\
        &+ \frac{\sigma^{\frac32}}{2} \left [ B_3 e^{ 2 i \omega t} + B_3^* e^{- 2 i \omega t}  + C_3 \right ] \\
        &+  \sigma^2 A^2 \sigma^{\frac12} \frac{(\sigma + 3 i \omega)(4 - i \omega)}{(\sigma^2 + 9 \omega^2)^2}
    \end{aligned}
\end{equation}
From Eqs. \eqref{B2C2} and \eqref{B3C3}, it is clear that among the three terms on the right-hand side of Eq. \eqref{combode9}, the first one is $O(\sigma^{\frac12})$ and the remaining ones are $O(\sigma^{-\frac12})$. Hence for $\sigma \gg 1$

\begin{equation}
    u_1 \approx \frac{10}{7} \frac{A^2}{\sigma^{\frac12}} \frac{1}{\sigma + 2 i \omega} e^{2 i \omega t}  - \frac{2A^2}{\sigma^{\frac12}}
    \label{uonetwo}
\end{equation}
and consequently, 

\begin{equation}
    B_1 \approx \frac{10}{7} \frac{A^2}{\sigma^{\frac12}} \frac{\sigma - 2 i \omega}{\sigma^2 + 4 \omega^2} \text{, } C_1 \approx - \frac{2A^2}{\sigma^{\frac12}}
    \label{B1C1}
\end{equation}

We return to Eq. \eqref{combode1} to extract the $e^{i \omega t}$ term from the right-hand side and note that the right-hand side of Eq. \eqref{combode1} which yields such a term is:
\begin{equation*}
    \begin{aligned}
        &-\frac{\sigma^{\frac52}}{2} (u_0 v_1 + u_1 v_0) + \frac{\sigma^{\frac32}}{2} (D+1) (u_0 v_1 + u_1 v_0) \\
        &- \frac{\sigma^2}{2} (D+2) (u_1 w_0 + u_0 w_1) - \frac{\sigma^2}{2} (u_1 w_0 + u_0 w_1) \\
        &+ \frac{\sigma^2}{4} \left [ (D+1)(D+2) + \sigma \right ](w_0 v_1 + w_1 v_0)
    \end{aligned}
\end{equation*}
The dominating term in the above expression for $\sigma \gg 1$ is the first term in which the coefficient of $e^{i \omega t}$ is $-\frac{1}{2} \sigma^{\frac52} A^3 \left [ \left ( - \frac{4}{\sigma^{\frac32}} \right ) + \left (- \frac{2}{\sigma^{\frac12}} \right ) \frac{2}{\sigma}  \right ] =  4 A^3 \sigma $. Using this approximation for the right-hand side of Eq. \eqref{combode1}, with the left-hand side obtained from Eq. \eqref{combode7}, the final amplitude equation becomes

\begin{equation}
    \label{finalamp}
    \left [ 4 \sigma - 2 i \omega_H \sigma  \right ] \frac{dA}{dt} = 8 \sigma (\Delta r) A - 4 A^3 \sigma + O \left (\frac1{\sigma} \right )
\end{equation}


\begin{thebibliography}{0}%
\makeatletter
\providecommand \@ifxundefined [1]{%
 \@ifx{#1\undefined}
}%
\providecommand \@ifnum [1]{%
 \ifnum #1\expandafter \@firstoftwo
 \else \expandafter \@secondoftwo
 \fi
}%
\providecommand \@ifx [1]{%
 \ifx #1\expandafter \@firstoftwo
 \else \expandafter \@secondoftwo
 \fi
}%
\providecommand \natexlab [1]{#1}%
\providecommand \enquote  [1]{``#1''}%
\providecommand \bibnamefont  [1]{#1}%
\providecommand \bibfnamefont [1]{#1}%
\providecommand \citenamefont [1]{#1}%
\providecommand \href@noop [0]{\@secondoftwo}%
\providecommand \href [0]{\begingroup \@sanitize@url \@href}%
\providecommand \@href[1]{\@@startlink{#1}\@@href}%
\providecommand \@@href[1]{\endgroup#1\@@endlink}%
\providecommand \@sanitize@url [0]{\catcode `\\12\catcode `\$12\catcode
  `\&12\catcode `\#12\catcode `\^12\catcode `\_12\catcode `\%12\relax}%
\providecommand \@@startlink[1]{}%
\providecommand \@@endlink[0]{}%
\providecommand \url  [0]{\begingroup\@sanitize@url \@url }%
\providecommand \@url [1]{\endgroup\@href {#1}{\urlprefix }}%
\providecommand \urlprefix  [0]{URL }%
\providecommand \Eprint [0]{\href }%
\providecommand \doibase [0]{https://doi.org/}%
\providecommand \selectlanguage [0]{\@gobble}%
\providecommand \bibinfo  [0]{\@secondoftwo}%
\providecommand \bibfield  [0]{\@secondoftwo}%
\providecommand \translation [1]{[#1]}%
\providecommand \BibitemOpen [0]{}%
\providecommand \bibitemStop [0]{}%
\providecommand \bibitemNoStop [0]{.\EOS\space}%
\providecommand \EOS [0]{\spacefactor3000\relax}%
\providecommand \BibitemShut  [1]{\csname bibitem#1\endcsname}%
\let\auto@bib@innerbib\@empty
\end{thebibliography}%


\begin{thebibliography}{30}
\bibitem{ref1} L. D. Landau and E. M. Lifshitz, ``Fluid mechanics, Vol. 6 of A course of Theoretical Physics,'' Pergamon Press, (1959).

\bibitem{ref2} D. Ruelle and F. Takens, ``On the nature of turbulence,'' Commun. Math. Phys. 20, 167 (1971) and 23, 343 (1971).

\bibitem{ref3} S. Newhouse, D. Rulle  and F. Takens, ``Occurrence of strange Axiom A attractors near quasi-periodic flows on $T^m$, $m\ge3$,'' Commun. Math. Phys. 64, 35 (1978).

\bibitem{ref4} E. N. Lorenz, ``Deterministic nonperiodic flow,'' J. Atmos. Sc. 20, 130 (1963).

\bibitem{ref5} See for e.g. S. H. Strogatz, ``Nonlinear dynamics and chaos,'' Westview Pres (2015).

\bibitem{ref6} M. J. Feigenbaum, ``Quantitative universality for a class of nonlinear transformations,''  J. Stat. Phys, 19, 25 (1978).

\bibitem{ref7} J. Guckenheimer, ``Sensitive dependence to initial conditions for one-dimensional maps,'' Comm. Math. Phys. 70 133 (1979).

\bibitem{ref8} M. J. Feigenbaum, ``The onset spectrum of turbulence,'' Phys. Lett. A 74, 375 (1979).

\bibitem{ref9} W. S.-Stupica, ``The behavior of nonlinear vibrating system,'' Kluwer Academic, London (1990).

\bibitem{ref10} A. Libchaber, C. Laroche and S. Fauve, ``Period doubling cascade in mercury, a quantitative measurement,'' J. Physique Lett. 43, 211 (1982).

\bibitem{ref11} S. Ramaswamy, ``The mechanics and statics of active matter,'' Ann. Rev. Condens. Matter Phys. 1, 323 (2010).

\bibitem{ref12} M. C. Marchetti, J. F. Joanny, S. Ramaswamy, T. B.Liverpool, and J. Prost, ``Hydrodynamics  of soft active matter,'' Rev. Mod. Phys. 85, 1143 (2013).

\bibitem{ref13} D. Saintillan and M. Shelley, ``Active suspensions and their nonlinear models,'' Comptes Rendus Physique, 14, 497 (2013).

\bibitem{ref14} M. Doi and S. F. Edwards, ``The theory of polymer dynamics,'' Oxford University Press, Oxford (1986).

\bibitem{ref15} S. M. Fielding, D. Marenduzzo, and M. E. Cates, ``Nonlinear dynamics and rheology of active fluids: Simulations in two dimensions,'' Phys. Rev. E 83, 041910 (2011).

\bibitem{ref16} H. Brand, H.Pleiner, and D. Svensek, ``Reversible and dissipative macroscopic contributions to the stress tensor: Active or passive?,'' Eur. Phys. J. E 37, 83 (2014).

\bibitem{ref17} D. Saintillan  and M. J. Shelley, ``Instabilities, pattern formation, and mixing in active suspensions,'' Phys. Fluids 20, 123304 (2008).

\bibitem{ref18} A. Tiribocchi, R. Wittkowski, D.Marenduzzo, and M.Cates, ``Active model H: Scalar active matter in a momentum-conserving fluid,'' Phys. Rev. Lett. 115,8 188302 (2015).

\bibitem{ref19} P. C. Hohenberg and B. I. Halperin, ``Theory of dynamic critical phenomena,'' Rev. Mod. Phys. 49, 435 (1977).

\bibitem{ref20} T. R. Kirkpatrick and J. K. Bhattacherjee, ``Driven active matter: Fluctuations and a hydrodynamic instability,'' Phys. Rev. Fluids 4, 024306 (2019).

\bibitem{ref21} N. M. Krylov and N. N. Bogolyubov, ``Introduction to Nonlinear Mechanics,'' Princeton University Press, Princeton (1947).

\bibitem{ref22} D. Smith, ``Singular Perturbation Theory'', Cambridge University Press, Cambridge (1985).

\bibitem{ref23} C. Sparrow, ``The Lorenz equations: bifurcations, chaos, and strange attractors,'' Springer-Verlag, New York (1982).

\bibitem{ref24} P. Glendinning and C. Sparrow, ``Local and global behavior near homoclinic orbits,'' J. Stat. Phy. 35, 645 (1984).

\bibitem{ref25} P. Gaspard and G. Nicolis, ``What can we learn from homoclinic orbits in chaotic dynamics?'' J. Stat. Phy. 31, 499 (1984).

\bibitem{ref26} J. B. McLaughlin and P. C. Martin, ``Transition to turbulence in a statistically stressed fluid system,'' Phys. Rev. A 12, 186-203 (1975).

\bibitem{ref27} D. Das, D. Banerjee and J. K. Bhattacharjee, ``Super-critical and sub-critical Hopf bifurcations in two and three dimensions,'' Nonlinear Dynamics 77, 169-184 (2014).

\bibitem{ref28} R. Chatterjee, N. Rana, R. Aditi Simha, P. Perlekar and S. Ramaswamy, ``Fluid flocks with inertia,'' arXiv 1907.03492 (2019).

\bibitem{ref29} R. Singh and M. E. Cates, ``Hydrodynamically interrupted droplet growth in scalar active matter,'' arXiv 1907.04819 (2019).

\bibitem{ref30} S. Shankar and M. Cristina Marchetti, ``Hydrodynamics of active defects: from order to chaos to defect ordering,'' arXiv 1907.02468 (2019).

\end{thebibliography}
 \end{document}